\documentclass[aps,prl,twocolumn,superscriptaddress,showpacs]{revtex4}

\usepackage{graphicx}
\usepackage{amsmath}
\usepackage{bm}

\renewcommand{\vec}[1]{\bm{#1}}

\begin{document}

\title{Field-induced disorder in a gapped spin system with non-magnetic impurities }

\author{H.-J. Mikeska}
\affiliation{Institut f\"ur Theoretische Physik, Universit\"at
Hannover, 
30167 Hannover, Germany}

\author{Asimkumar Ghosh}
\affiliation{Institut f\"ur Theoretische Physik, Universit\"at
Hannover, 
30167 Hannover, Germany}
\affiliation{Department of Physics, Scottish Church College,
Kolkata 700006, India.}

\author{A. K. Kolezhuk}
\affiliation{Institut f\"ur Theoretische Physik, Universit\"at
Hannover, 
30167 Hannover, Germany}
\affiliation{Institute of Magnetism, National Academy of Sciences and
Ministry of Science and Education, 
03142 Kiev, Ukraine }

\begin{abstract}
We study $S=\frac{1}{2}$ dimers which are weakly coupled by
three-dimensional antiferromagnetic interactions $J'$. It is commonly
known that doping with non-magnetic impurities immediately induces a
long-range N\'eel order. We show that application of an external
magnetic field $H$ may drive the system back to a disordered phase.
We discuss the zero temperature phase diagram in the $(H,J')$ plane and
we propose suggestions for experiments.
\end{abstract}


\maketitle

In the last decade, much attention has been paid to the physics of
gapped spin systems doped with non-magnetic impurities. Typical
examples include $S=\frac{1}{2}$ ladders
\cite{SigristFurusaki96,Nagaosa+96,IinoImada96,MikNS97,Azuma+97},
$S=\frac{1}{2}$ dimerized chains \cite{Martin+97}, Haldane spin chains
\cite{ShenderKivelson91,Uchiyama+99}, various two-dimensionally
coupled systems \cite{Martins+97,ImadaIino97,Wessel+01,Yasuda+01} and
systems of $S=\frac{1}{2}$ dimers weakly coupled by three-dimensional
(3D) interaction \cite{Oosawa+02,Oosawa+03}. It was shown
\cite{SigristFurusaki96,MikNS97,Martins+97} that the presence of
impurities induces the formation of a gapless continuum of low-lying
states within the gap, which leads to long-range N\'eel ordering in
the presence of arbitrarily small 3D interactions.

On the other hand, considerable interest has been devoted to the
behavior of gapped spin systems in strong magnetic fields. When the
field is strong enough to close the spectral gap, the system enters a
new phase. In absence of anisotropy, this new phase is critical in the
purely one-dimensional (1D) case and exhibits long-range order in the
presence of a weak 3D coupling. Experimentally, field-induced ordering
has been studied for several substances, one of the better understood
examples being the dimer material $\rm TlCuCl_{3}$
\cite{Nikuni+00,Tanaka+01,Matsumoto+02,Ruegg+03}.

Consider an interacting 3D system of spin-$\frac{1}{2}$ dimers
described by the following Hamiltonian:
\begin{eqnarray} 
\label{ham-3Ddimer} 
\mathcal{H}&=&J\sum_{\vec{r}} \vec{S}_{\vec{r},1}\cdot\vec{S}_{\vec{r},2}
-H\sum_{\vec{r}\sigma}S_{\vec{r},\sigma}^{z}
\nonumber\\
&+&J'\sum_{\langle \vec{r}\vec{r'}\rangle}
(\vec{S}_{\vec{r},1}\cdot\vec{S}_{\vec{r'},2}+
\vec{S}_{\vec{r},2}\cdot\vec{S}_{\vec{r'},1}), 
\end{eqnarray}
where $J>0$ and $J'>0$ are the intra- and interdimer exchange
couplings, respectively, and $H$ is the external magnetic field
directed along the $z$ axis. The vector $\vec{r}$ labels the dimers
located at the sites of a lattice with coordination number $Z$, and
$\langle ..\rangle$ denotes summation only over neighboring dimers. In
(\ref{ham-3Ddimer}) we have assumed that the couplings are
unfrustrated so that each spin $\vec{S}_{\vec{r},\sigma}$ can be
classified as belonging to one of the two sublattices (respectively,
$\sigma=1$ or $\sigma=2$).

In the absence of impurities and external field and for sufficiently
weak interdimer coupling $J'\ll J$ the system has a singlet ground
state without any magnetic order, and a finite gap $\Delta\sim J$ to
the lowest excitation which is a triplet; known examples of materials
exhibiting this type of ground state are $\rm KCuCl_{3}$
\cite{Tanaka+96} and $\rm TlCuCl_{3}$ \cite{Takatsu+97}. If the 3D
coupling $J'$ exceeds some critical value (typically of the order of
$J/Z$), long-range N\'eel order appears; this situation is realized
e.g. in $\rm NH_{4}CuCl_{3}$ \cite{Shiramura+98}. An external
magnetic field closes the gap at the critical field $H = H_{c} =
\Delta$, inducing a phase transition. At $H>H_{c}$ a finite
magnetization along the field direction appears, accompanied by a
staggered order in the plane perpendicular to the field; the U(1)
symmetry in the $xy$ plane is spontaneously broken.  This transition
may be viewed as the Bose-Einstein condensation of magnons
\cite{Nikuni+00}.

The presence of non-magnetic impurities drastically changes the above
picture \cite{SigristFurusaki96,MikNS97,ImadaIino97}. Impurities
generate unpaired spins, which develop an effective interaction with
each other mediated by the intact dimers between them. According to
the Lieb-Schulz-Mattis theorem, the interaction between two unpaired
spins is antiferromagnetic if they belong to the same sublattice and
ferromagnetic otherwise, so in total the unpaired spins encourage
 an antiferromagnetic ordering which at $T=0$ occurs
at any arbitrarily small concentration of impurities $c$. Unpaired
spins thus form a separate subsystem with the energy scale for its
dynamics set by the average effective interaction $\widetilde{J} \ll
\Delta$. The excitation spectrum of the impure system is determined by
the gapless continuum of the low-lying states resulting from the
ordered unpaired-spin subsystem; this continuum coexists with the
states above the spin gap $\Delta$ of the pure system which survive
with reduced weight.

So, taken \emph{separately}, both doping with non-magnetic impurities
and application of an external field tend to induce ordering in a
gapped spin system. The aim of the present Letter is to show that
application of an external field to a doped system may drive it back
into the disordered phase, causing reentrant behavior of the N\'eel
order as a function of $H$. The qualitative form of the $T=0$ phase
diagram in the $(H,J')$ space will be obtained from a mean-field
treatment of the interaction between the dimer and unpaired spin
subsystems.

\emph{Dynamics of an intact dimer system.--}
To derive the effective interaction between two unpaired spins, we
have to consider the dynamics of the pure dimer system first. For this
purpose we use the dimer field theory \cite{K96} which may be viewed
as a continuum version of the bond boson operator approach
\cite{Matsumoto+02}. The quantum state is formulated as the product of
dimer states of the form
\begin{equation}
|\psi\rangle=(1-A^2-B^2)^{1/2}
|s\rangle+\sum_j (A_{j}+iB_{j})|t_j\rangle,
\label{dimer-wf}
\end{equation}
where $|s\rangle$ denotes the singlet state and $|t_j\rangle$, $j=x,y,z$ are the
three triplet states of the spin dimer in the Cartesian basis.
The vectors $\vec{A}$, $\vec{B}$  are 
connected with the magnetization
$\vec{M}=\langle\vec{S}_1+\vec{S}_2\rangle$ and sublattice magnetization
$\vec{L}=\langle\vec{S}_1-\vec{S}_2\rangle$ of the spin dimer as follows:
\begin{equation}
\vec{M}=2(\vec{A}\times\vec{B})\;,\quad
\vec{L}=2(1-A^2-B^2)^{1/2}\vec{A}\;.
\label{ML}
\end{equation}
The Lagrangian density (per dimer) can be written as
\begin{eqnarray}  
\label{Lagr} 
&& \mathcal{L}=2\hbar \vec{B}\cdot\frac{\partial\vec{A}}{\partial t} -E_{\rm
  dim} -\frac{ZJ'}{4}\big\{ (\nabla \vec{L})^{2}+ (\nabla
  \vec{M})^{2} \big\} \nonumber \\
&& E_{\rm dim}=J(A^{2}+B^{2}) +
  \frac{ZJ'}{4}(\vec{M}^{2}-\vec{L}^{2})-HM_{z}, 
\end{eqnarray}
where we use the notation $(\nabla \vec{A})^{2}\equiv
\sum_{i}(\partial\vec{A}/\partial x_{i})^{2}$.
Assuming $A,B \ll 1$, at the quadratic level
one easily obtains the dispersion of a magnon with $S^{z}=\mu$ as 
$\varepsilon(\vec{k})=\Delta(1+ k^{2}\xi^{2})^{1/2} -\mu H$, where
\begin{equation} 
\label{vDelta}
\Delta=[J(J-ZJ')]^{1/2},\quad \xi=\frac{v}{\Delta},\quad
v=\Big(\frac{ZJJ'}{2}\Big)^{1/2}.
\end{equation}
Here $\xi$ is the spin correlation length, $\xi\ll 1$ if the 3D
interaction is small and $ZJ'\ll J$ (the lattice constant is set to
unity).

\emph{Interactions between unpaired spins.--}
The effective interaction between two unpaired spins can be estimated
\cite{SigristFurusaki96} in second-order perturbation theory,
\begin{equation} 
\label{Jeff1} 
J_{\rm eff}= \mathcal{K}\int d^{3}k 
\frac{(ZJ')^{2}}{\varepsilon(k)}e^{i\vec{k}\cdot\vec{r}} 
 =\frac{4\pi\mathcal{K} (ZJ')^{2}}{r v \xi} K_{1}\Big(\frac{r}{\xi}\Big),
\end{equation}
where $\mathcal{K}$ is an unknown constant of the order of unity,
determined by the matrix elements of the perturbation, $\vec{r}$ is
the vector connecting the two spins, $K_{1}$ is the MacDonald function,
and we omit the oscillating sign depending on whether or not the spins
belong to the same sublattice; we have used the $H=0$ expression for
the magnon energy since, as will be clear below, we are interested in
fields which are small comparing to the gap $\Delta$. If $r=1$, there
is an additional direct (first-order in the perturbation $J'$)
interaction between the spins. Since $\xi\ll 1$, one can write down
the effective interaction as
\begin{equation} 
\label{Jeff2} 
J_{\rm eff}(r)=J_{0}\delta_{r,1}+  J_{1} r^{-3/2} e^{-r/\xi},
\end{equation}
where $J_{0} \sim J'$ and $J_{1} = (2\sqrt{2}\pi)^{3/2}\mathcal{K}
J'(Z^{5}J'/J)^{1/4}$.  Since $J_{\rm eff}$ is decaying very fast with
increasing $r$, it is sufficient to take into account interactions
only between those unpaired spins which are nearest neighbors. For a
given impurity concentration $c$, the distribution of distances $R$
from an impurity to its nearest neighbor can be estimated in the
continuum approximation (i.e. neglecting the presence of the lattice)
as
\begin{equation} 
\label{pR} 
p(R)=c\, \exp\{-\frac{4\pi c}{3}R^{3}\},\quad \int p(R) 4\pi R^{2}\,dR=1.
\end{equation}
The distribution of effective exchange couplings $\widetilde{J}$
between the neighboring unpaired spins is given by
\[
P(\widetilde{J})=\sum_{\vec{R}} p(R)\, \delta(\widetilde{J}-J_{\rm eff}(R)).
\]
Replacing the sum by an integral and substituting the exact value of
$p(1)=Zc$, one obtains
\begin{eqnarray} 
\label{pJ}
P(\widetilde{J})&=&Zc\delta(\widetilde{J}-J_{0}) +P_{\rm reg}(\widetilde{J}),\\
P_{\rm reg}(\widetilde{J})&=&\Theta(J_{1}e^{-1/\xi} -\widetilde{J})\,
\frac{4\pi c\xi}{\widetilde{J}}\, R_{1}^{2}\,\exp\{ -4\pi c R_{1}^{3}/3 \},
\nonumber
\end{eqnarray}
where $R_{1}\simeq \xi \ln(J_{1}/\widetilde{J})$ and $\Theta$ is the
Heaviside function;  note that $P(\widetilde{J})$ has an
upper cutoff.  The peak of the regular part $P_{\rm reg}$ occurs at 
$\widetilde{J}=\widetilde{J}_{\rm peak}\simeq
J_{1}\exp\{ -(4\pi c\xi^{3})^{-1/2}\}$, and the ``regular average'' is
\begin{equation} 
\label{Jav} 
\langle\widetilde{J}\rangle_{\rm reg}=\int J P_{\rm reg}(J) \, dJ 
\simeq 4\pi c\xi J_{1}e^{-1/\xi} \gg \widetilde{J}_{\rm peak}.
\end{equation}

\begin{figure}[tb]
\includegraphics[width=70mm]{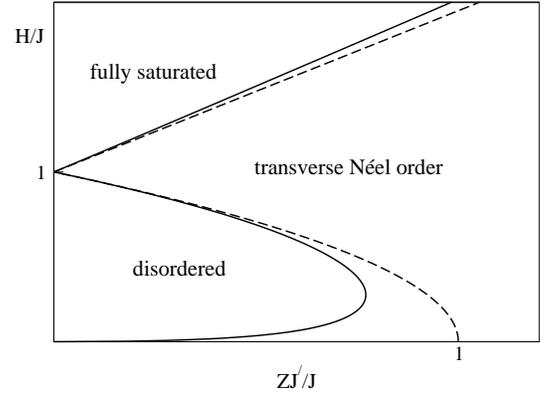}
\caption{\label{fig:imp-phd} A schematic mean-field $T=0$ phase
diagram of the model (\protect\ref{ham-3Ddimer}).The dashed line
corresponds to a pure system and the solid line to a finite
concentration of impurities. The boundaries of the disordered and
saturated phases are determined by (\protect\ref{stab1}) and
(\protect\ref{Hs}), respectively.}
\end{figure}

\emph{The interacting system of unpaired spins and intact dimers.--} 
We go on to study the static properties 
(e.g. the sublattice
magnetization) 
of the impure system in a magnetic field in a
mean-field approximation by minimizing the full energy $E=E_{\rm
dim}+E_{\rm unp}+E_{\rm int}$. Here, $E_{\rm dim}$ is the energy of
the intact dimers, Eq.~(\ref{Lagr}), $E_{\rm unp}$ is the energy of
the subsystem of unpaired spins and $E_{\rm int}$ the interaction
energy between the unpaired spins and the intact dimer subsystem. 
The unpaired spin subsystem is ordered, and we therefore describe it
as a two-sublattice antiferromagnet.  Denoting the angle between the
sublattice magnetization and the $z$ axis as $\theta$ and the average
spin length of an unpaired spin as $s$ , we write the interaction
energy (per dimer) as
\begin{equation} 
\label{eint} 
E_{\rm int}= c ZJ's (M_{z}\cos\theta -L_{x}\sin\theta),
\end{equation}
where the $x$ axis is chosen along the direction of the staggered
order. In the self-energy $E_{\rm unp}$ of the unpaired spin
subsystem we include, for each unpaired spin, the interaction with
nearest unpaired neighbors only; this gives 
\begin{equation} 
\label{eunp} 
E_{\rm unp}= 2c (-\frac{1}{2}Z_{i}\widetilde{J}s^{2}\sin^{2}\theta 
             -Hs\cos\theta). 
\end{equation}
Here $Z_{i}$ is some number which has the sense of an average number
of ``nearest neighbors'' for a fictitious random lattice formed by
unpaired spins, and the energy is again taken per dimer, hence the
factor $2c$ in front. The number $Z_{i}$ can be safely put to $1$, as
the following estimate shows: We would like to know how many unpaired
spins in the closest vicinity of a given one (which we put at the
origin) can have ``almost the same''effective interaction $J_{\rm
eff}$ with it. Since $J_{\rm eff}$ decays as $e^{-r/\xi}$, we have to
count impurities in a spheric layer with some radius $R$ (the distance
to the closest impurity, on average $R\sim c^{-1/3}$) and thickness
$h\sim\xi$. Thus the average number of other unpaired spins which are
at the same distance from the origin as the closest one is $4\pi
R^{2}\xi c \ll 1$, so $Z_{i}\simeq 1+4\pi c R^{2}\xi\simeq 1$.

We now derive from the mean-field approach the static properties for
fixed effective interaction $\widetilde{J}$ and afterwards perform an
averaging over $\widetilde{J}$.  It is convenient to parameterize the
vectors $\vec{A}$ and $\vec{B}$ as follows:
\[
\vec{A}=\{ \sin\alpha\cos\gamma,0,0\},\quad
\vec{B}=\{ 0,\sin\alpha\sin\gamma,0\}.
\]
It is easy to see that $\alpha=0$, $\theta=0$ is always an extremum of
$E$ describing a state with a fully polarized unpaired spin subsystem
and a non-magnetic (singlet) dimer subsystem. A trivial analysis yields
the following equation for the stability boundary of this solution:
\begin{eqnarray} 
\label{stab1} 
&&(H^{2}-J^{2})(H-\widetilde{J}s)  +cs(ZJ')^{2}(J+csH-cs^{2}\widetilde{J}) \nonumber\\
&& \qquad\qquad\qquad + ZJ'(H-\widetilde{J}s)(J-2csH) =0.
\end{eqnarray}
Another obvious solution is the saturated state with all spins fully
polarized, which corresponds to $\theta=0$, $\alpha=\pi/2$,
$\gamma=\pi/4$. This state becomes unstable if the field drops below
the saturation value
\begin{equation} 
\label{Hs} 
H_{s}=J+ZJ'(1+cs).
\end{equation}
At fixed $J'$ the equation (\ref{stab1}) has two solutions for $H$. Up
to the second order in $J'$, $\widetilde{J}$ and to the first order in $c$ the
lower and upper critical fields are given by
\begin{eqnarray} 
H_{c1}&\simeq& \widetilde{J}s +cs(ZJ')^{2}/J,  \label{Hc1} \\
H_{c2} &\simeq& J- ZJ'(1/2-cs) -(ZJ')^{2}/(4J). \label{Hc2}
\end{eqnarray}
It is easy to obtain the behavior of $\alpha$, $\theta$, $\gamma$ in
the vicinity of the lower critical field $H_{c1}$:
\begin{eqnarray} 
\label{atg}
&& \theta\simeq [2(1-H/H_{c1})]^{1/2},\quad 
\alpha\simeq csZJ' \theta\cos\gamma,\nonumber\\ 
&&\gamma\simeq (\widetilde{J}-cZJ')s/2. 
\end{eqnarray}
The total staggered magnetization per dimer in the vicinity of
$H_{c1}$ vanishes as a square root:
\begin{equation} 
\label{mst} 
m_{\rm st} \simeq cs(1+ZJ'/J)[8(1-H/H_{c1})]^{1/2},
\end{equation}
where the term proportional to $J'$ describes the contribution of
intact dimers. Thus, the following physical picture emerges: At zero
external field, the presence of impurities induces a finite staggered
order, $\alpha\not=0$. When the external field $H$ is switched on, it
induces canting of spins in the field direction, i.e. $\theta$ starts
to deviate from $\pi/2$. However, with increasing $H$ the canting
angle $\theta$ moves towards $0$, diminishing the effective staggered
field from the unpaired spins acting on the dimer subsystem. Unpaired
spins become saturated at $H=H_{c1}$ ($\theta$ becomes zero), and
simultaneously the staggered order in the dimer subsystem vanishes
($\alpha$ becomes zero). The system enters a disordered phase where
the dimers are in a singlet state and unpaired spins are fully
magnetized. The resulting phase diagram is sketched in Fig.\
\ref{fig:imp-phd}.

\begin{figure}[tb]
\includegraphics[width=70mm]{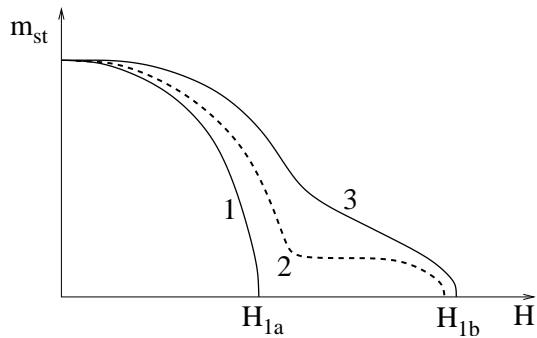}
\caption{\label{fig:mst} Schematic behavior of the staggered magnetization as
function of the field $H$. Curves 1, 2, and 3 correspond to the regimes
$\alpha\ll x_{1}$, $x_{1}\ll \alpha \ll x_{2}$, and $\alpha\gg x_{2}$,
respectively, $\alpha=ZJ'/(2J)$ and $x_{1,2}$ as in
(\protect\ref{ineq1}), (\protect\ref{ineq2}).  }
\end{figure}

As a byproduct of this description, one can study how the high-energy
spectrum (above the gap $\Delta$) is affected by presence of
impurities. One may replace $E_{\rm dim}$ in the Lagrangian
(\ref{Lagr}) by the full energy $E$ and assume that $\theta$ is frozen
at its equilibrium value. This amounts to neglecting the dynamics of
the unpaired spin subsystem which occurs at a much lower energy scale
$\widetilde{J}$, justifying to take into account only the static part
described by a fixed $\theta$. In that way, one obtains that the
triplet gap splits due to the appearance of the staggered order in the
$x$ direction; at $H=0$, the gaps are related to the order parameter
$A_{0}\propto m_{\rm st}$ as follows:
\begin{equation} 
\label{split} 
\Delta_{y,z}=(\Delta^{2}+4v^{2}A_{0}^{2})^{1/2}, \quad
\Delta_{x}=(\Delta^{2}+12v^{2}A_{0}^{2})^{1/2}.
\end{equation}
For $A_{0}\ll 1$, the square of the order parameter determining the
intensity of the Bragg peak is approximately linearly related to the
gap energy.  This proportionality between the change in the gap energy
and the intensity of the Bragg peak was observed experimentally in
$\rm TlCuCl_{3}$ doped with $\rm Mg$ \cite{Oosawa+03} by varying the
temperature. Our calculation is for $T=0$, but we believe that this
relation continues to be valid for finite temperature as well.

\emph{Averaging over the impurity distribution.--}
As final step we have to average the staggered magnetization
(\ref{mst}) with $H_{c1}$ given by (\ref{Hc1}), using the distribution
function (\ref{pJ}). Formally, the field $H_{0}$ at which $m_{\rm st}$
completely disappears is determined by the upper cutoff of
$P(\widetilde{J})$. However, the singular ($\delta$-function) part of
$P(\widetilde{J})$ has to be neglected for two reasons: (a) it has a
small weight of $Zc$, so that the corresponding contribution to
$m_{\rm st}$ is of the order of $c^{2}$; (b) it describes unpaired
spins located at neighboring sites, and they will have a strong
tendency to build a singlet state since their interaction with each
other is much stronger than with the rest of the unpaired spin
``network''.  The regular part $P_{\rm reg}$ of the distribution
carries the main weight $1-Zc$ and corresponds to much smaller
$\widetilde{J}$ with the cutoff at
$\widetilde{J}=J_{1}e^{-1/\xi}$. Its contribution to $m_{\rm st}$
vanishes above the characteristic field
\begin{equation} 
\label{H1} 
H_{1}=sJ_{1}e^{-1/\xi} + cs(ZJ')^{2}/J  
\end{equation}
One has to distinguish two different regimes, depending on the strength of the
interdimer interaction. For a very small 
interdimer coupling 
\begin{equation} 
\label{ineq1} 
\frac{ZJ'}{J} \ll 2 x_{1}, \quad \text{where}\quad x_{1}^{3/4} e^{1/\sqrt{x_{1}}}
= (2\pi)^{3/2}\frac{\mathcal{K}}{c},
\end{equation}
the
second term in (\ref{H1}) is always leading, and 
\[
H_{1}\simeq H_{1a}=
cs(ZJ')^{2}/J,
\]
with $m_{\rm st}$ behaving  as a pure square root $\sqrt{H_{1a}-H}$.
For stronger $J'$ 
the inequality  (\ref{ineq1}) is violated, then the
first term in (\ref{H1}) becomes leading, and 
\[
H_{1}\simeq H_{1b}=sJ_{1} e^{-1/\xi}.
\]
The behavior of $m_{\rm st}$ for $H < H_{1}$ depends again on the
interdimer interaction strength. If $\langle \widetilde{J}
\rangle_{\rm reg} \ll c(ZJ')^{2}/J$, which corresponds to weak
interdimer coupling $ZJ'/J\ll 2x_{2}$ with $x_{2}\gg x_{1}$ given by
\begin{equation} 
\label{ineq2} 
x_{2}^{1/4}\exp\big\{x_{2}^{-1/2}\big\} = 2(2\pi)^{5/2}\mathcal{K},
\end{equation}
then the main contribution to the staggered magnetization $m_{\rm st}$
has an overall $\sqrt{H_{1a} -H}$ behavior, and additionally there is
a weak shoulder extending to a higher field $H=H_{1b}$. The shoulder
strength grows with increasing $J'$ and for $2x_{2}\ll ZJ'/J \ll 1$
merges with the main part, as sketched in Fig.\ \ref{fig:mst}.

In summary, we have shown that in a system of weakly coupled spin
dimers doped with non-magnetic impurities, the application of an
external field $H$ may lead to a reentrant behavior of long-range
antiferromagnetic order as a function of field. From the phase diagram
of Fig.\ \ref{fig:imp-phd} it is clear that this phenomenon is most
likely to occur in materials with a sufficiently small ratio of inter-
to intradimer exchange. Therefore, among known materials, $\rm
KCuCl_{3}$ is the most promising candidate.

\emph{Acknowledgments.--} 
We are grateful to  K.~Kakurai, H.~Tanaka and A.~Oosawa for useful comments.  
This work is supported in part by the grant I/75895 from the
Volkswagen-Stiftung.


\begin{thebibliography}{50}

\bibitem{SigristFurusaki96} M.~Sigrist and A.~Furusaki,  
                J. Phys. Soc. Jpn {\bf 65}, 2385 (1996).

\bibitem{Nagaosa+96} N. Nagaosa, A. Furusaki, M. Sigrist, and H. Fukuyama,
  J. Phys. Soc. Jpn. {\bf 65}, 3724 (1996).

\bibitem{IinoImada96} Y. Iino and M. Imada, J. Phys. Soc. Jpn. {\bf 65}, 
  3728 (1996).

\bibitem{MikNS97} H.-J.~Mikeska, U.~Neugebauer, and U.~Schollw\"ock, 
                Phys. Rev. B {\bf 55}, 2955 (1997). 

\bibitem{Azuma+97} M. Azuma, Y. Fujishiro, M. Takano, M. Nohara, and
H. Takagi, Phys. Rev. B {\bf 55}, R8658 (1997).


\bibitem{Martin+97} M. C. Martin, M. Hase, K. Hirota, 
G. Shirane, Y. Sasago, N. Koide, and K. Uchinokura, Phys. Rev. B. {\bf 56}, 3173 (1997).


\bibitem{ShenderKivelson91} E. F. Shender and S. A. Kivelson,
  Phys. Rev. Lett. {\bf 66}, 2384 (1991).

\bibitem{Uchiyama+99} Y. Uchiyama, Y. Sasago, I. Tsukada, K. Uchinokura,
  A. Zheludev, T. Hayashi, N. Miura, and P. B\"oni, Phys. Rev. Lett. {\bf 83},
  632 (1999).

\bibitem{Martins+97} G. B. Martins, M. Laukamp, J. Riera, and E. Dagotto,
  Phys. Rev. Lett. {\bf 78}, 3563 (1997).

\bibitem{ImadaIino97}  M. Imada and Y. Iino, J. Phys. Soc. Jpn. {\bf 66}, 568 (1997).

\bibitem{Wessel+01}     S. Wessel, B. Normand, M. Sigrist, and S. Haas,
  Phys. Rev. Lett. {\bf 86}, 1086 (2001).

\bibitem{Yasuda+01} C. Yasuda, S. Todo, M. Matsumoto, and H. Takayama,
  Phys. Rev. B {\bf 64}, 092405 (2001). 


\bibitem{Oosawa+02} A.~Oosawa, T.~Ono, and H.~Tanaka,
                Phys. Rev. B {\bf 66}, 020405(R) (2002).

\bibitem{Oosawa+03} A.~Oosawa, M.~Fujisawa, K.~Kakurai, and H.~Tanaka,  
                Phys. Rev. B {\bf 67}, 184424 (2003).

\bibitem{Nikuni+00} T. Nikuni, M. Oshikawa, A. Oosawa, and H. Tanaka,
  Phys. Rev. Lett. {\bf 84}, 5868 (2000).

\bibitem{Tanaka+01} H. Tanaka,  A. Oosawa, T. Kato, H. Uekusa, Y. Ohashi,
  K. Kakurai,  and A. Hoser, J. Phys. Soc. Jpn. {\bf 70}, 939 (2001).


\bibitem{Matsumoto+02} M.~Matsumoto, B.~Normand, T.M.~Rice and M.~Sigrist, 
                Phys. Rev. Lett. {\bf 89}, 077203 (2002).
                
\bibitem{Ruegg+03} Ch. R\"uegg, N. Cavadini,
  A. Furrer, H.-U. G\"udel, K.~Kr\"amer, H.~Mutka, A.~Wildes, K. Habicht, 
  and P. Vorderwisch, Nature \textbf{423}, 62 (2003).

\bibitem{Tanaka+96} H. Tanaka, K. Takatsu, W. Shiramura, and T. Ono,
  J. Phys. Soc. Jpn. {\bf 65}, 1945 (1996).

\bibitem{Takatsu+97} K. Takatsu, W. Shiramura, and H. Tanaka,
  J. Phys. Soc. Jpn. {\bf 66}, 1611 (1997).

\bibitem{Shiramura+98} W. Shiramura, K. Takatsu, B. Kurniawan,
H. Tanaka, H. Uekusa, Y. Ohashi, K. Takizawa, H. Mitamura, and
T. Goto: J. Phys. Soc. Jpn. \textbf{67}, 1548 (1998).

\bibitem{K96} A.~K. Kolezhuk: Phys. Rev. B \textbf{53}, 318 (1996).


\end{thebibliography}
\end{document}